\documentclass[twocolumn,showpacs,prc,amsmath,amssymb,nofootinbib]{revtex4}

\usepackage{graphicx}
\usepackage{dcolumn}
\usepackage{bm}
\usepackage{lscape}
\usepackage{mathrsfs}
\usepackage{color}

\begin{document}


\title{Channel coupling effect and important role of imaginary part of coupling potential\\ for high-energy heavy-ion scatterings}

\author{T.~Furumoto}
\email{furumoto@ichinoseki.ac.jp}
\altaffiliation[Present address: ]{Ichinoseki National College of Technology, Ichinoseki, Iwate 021-8511, Japan}
\affiliation{RIKEN Nishina Center, Wako, Saitama 351-0198, Japan}

\author{Y.~Sakuragi}%
\email{sakuragi@sci.osaka-cu.ac.jp}
\affiliation{Department of Physics, Osaka City University, Osaka 558-8585, Japan}
\affiliation{RIKEN Nishina Center, Wako, Saitama 351-0198, Japan}

%

\date{\today}

\begin{abstract}
The recent works by the present authors and their collaborator predicted that the real part of heavy-ion optical potentials changes its character from attraction to repulsion around the incident energy per nucleon $E/A =$ 200 -- 300 MeV on the basis of the complex $G$-matrix interaction and the double-folding model (DFM) and revealed that the three-body force plays an important role there.
In the present paper, we have analyzed the energy dependence of the coupling effect with the microscopic coupled channel (MCC) method and its relation to the elastic and inelastic-scattering angular distributions in detail in the case of the $^{12}$C + $^{12}$C system in the energy range of $E/A=$ 100 -- 400 MeV. 
The large channel coupling effect is clearly seen in the elastic cross section although the incident energies are enough high.
The dynamical polarization potential (DPP) is derived to investigate the channel coupling effect.
Moreover, we analyze the effect of the imaginary part of the coupling potential on elastic and inelastic cross sections.
\end{abstract}


\maketitle

\section{Introduction}
The collective excitation of nuclei is known to play an important role in heavy-ion (HI) reactions.
The strong coupling among the ground and low-lying collective states of colliding nuclei requires a non-perturbative treatment to properly account for the coupling effects on the elastic and inelastic scatterings. 
The coupled-channel (CC) method is one of the most reliable and established reaction theory to study the role of nuclear excitations in HI reactions and to extract nuclear-structure information through the CC analyses of the experimental data~\cite{DNR}. 

In the conventional CC calculations, the phenomenological optical potentials are used for constructing the diagonal and coupling potentials.
In the case of collective excitations, the form factors of the coupling potentials are given by the derivative forms of the optical potential (in the case of vibrational excitations) or by the deviation of the deformed optical potential from the spherical one (in the case of rotational excitations).
The strengths of the coupling potentials are determined so as to reproduce the known electric transition rates such as the $B$(E$\lambda$) values, if available. Otherwise, they are treated as the free parameters that are chosen so that the CC calculation reproduces the experimental data of the elastic and inelastic scattering~\cite{MIK71,MAT72,RAY78,COO86,REB94,ELE08}.

Despite the successful applications of the conventional CC method for various HI reactions, serious problems have been pointed out frequently in connection with large ambiguities in the shape and strength of the phenomenological optical potentials for HI systems~\cite{DNR,BRAN97}. 
The ambiguity of the optical potential adopted in the CC calculation leads to a serious difficulty in the proper evaluation of the channel-coupling effects as well as unknown nuclear-structure information such as the deformation lengths and electric transition rates.
In order to overcome these difficulties, microscopic coupled-channels (MCC)~\cite{DNR,SAK83,SAK86,KHO00} method have been proposed on the basis of microscopic optical potential models~\cite{SAT79}. 
In the MCC method, the diagonal and coupling potentials used in the CC calculations are constructed by the double-folding model (DFM) with the use of an effective nucleon-nucleon ($NN$) interaction.

On the early stage of the MCC studies of HI reactions, the effective $NN$ interactions called the M3Y interaction~\cite{M3Y77} or its density-dependent version called DDM3Y~\cite{KOB84,FARID85} (including its modified versions) have been used in constructing the diagonal and coupling potentials~\cite{SAT79,SAK83,SAK86,SAK87,SYKT88,KHO00,KAT02}. 
These interactions, especially the density-dependent versions, have been prove to give a good account of the strength and shape of the internuclear potentials.
However, all these effective $NN$ interactions have real part only and, therefore, one has to add an phenomenological imaginary part by hand to the diagonal and coupling potentials obtained by the DFM calculations with the real $NN$ interactions, which makes the results of CC calculations still ambiguous.
It is of particular importance to note that 
the channel-coupling effects largely depend on the real to imaginary ratio of the coupling potentials~\cite{SAK86,SAK87,KHO00}.

Recently, several types of microscopic interaction models that predict complex optical potentials for composite projectiles have been 
proposed and applied to the analyses of elastic and inelastic scattering. 
One is the complex DFM~\cite{HOG89,TRA00,FUR06} with the use of the Jeukenne-Lejeune-Mahaux (JLM) interaction~\cite{JLM77,JLM98}.
The JLM interaction is a very simple complex effective $NN$ interaction and easy to handle in the folding calculations and, hence, has widely been used in the nucleon-nucleus systems. 
Another widely-used interaction model is the S\~{a}o Paulo potential (SPP)~\cite{CHA98,SPP03,SPP09} that is the DFM potential multiplied by the local-velocity-dependent Pauli non-locality correction factor.
Both models still suffer from uncertainty originating from either existence of the free parameters or the lack of theoretical foundation of the models and assumptions.

The latest interaction model based on the complex $G$-matrix interaction~\cite{CEG07,FUR09} is the most fundamental microscopic model for complex optical potential that has been successfully applied to proton-nucleus~\cite{CEG07} and nucleus-nucleus~\cite{FUR09,FUR09R,FUR10,FUR12} elastic scattering over the wide range of incident energies.
In this model, a new type of complex effective $NN$ interaction called CEG07~\cite{CEG07} was constructed on the basis of the Breuckner $G$-matrix theory and the CEG07 interaction is doubly folded with the nucleon density distributions of the colliding nuclei giving a complex optical potential for the HI system. 
Because of its reliable microscopic foundation, it is interesting to apply the present complex interaction model to the coupled-channel calculations of the HI reactions.

It is rather straightforward to generalize the successful microscopic theory for complex HI optical potential to the study of inelastic scattering of HI system that excites low-lying collective excited states. 
Namely, it is just to replace the real effective $NN$ interaction (such as the DDM3Y one) by the complex one (CEG07) in the DFM calculation of the diagonal and coupling potentials within the MCC framework.
This kind of MCC method based on the complex $NN$ interaction was first applied to the elastic and inelastic scattering of $^{16}$O~+~$^{16}$O system at medium energies~\cite{TAK10}. 

Here, it should be noted that the microscopic HI optical potentials predicted by the DFM with CEG07 shows a characteristic energy dependence. The real part of the HI optical potential becomes shallower as the increase of the incident energy and changes its sign from negative (attractive) to positive (repulsive) at the incident energy per nucleon ($E/A$) around 300 MeV region, whereas the imaginary part of the optical potential gradually increase with the increase of the incident energy~\cite{FUR10}.
Although the precise energy region where the attractive to repulsive transition occurs is still to be examined through experimental confirmation~\cite{GUO12}, there is no doubt that the real to imaginary ratio of the optical potentials must drastically change in such medium to high energy region. 

This kind of characteristic behavior of the microscopic optical potential will manifest itself also in the complex coupling potentials calculated with the CEG07 interaction within the MCC framework.
In the present paper, we study the energy dependence of the real and imaginary parts of the coupling potentials derived from the CEG07 interaction and investigate its relation to the channel-coupling effects on the elastic and inelastic scattering of the  $^{12}$C~+~$^{12}$C system in the MCC framework.
Particular attention will be paid to the characteristic energy dependence of the so-called dynamical polarization potential (DPP) and its relation to the energy dependence of the real to imaginary ratio of the coupling potential predicted by the microscopic interaction model with the CEG07 interaction.

\section{Microscopic Coupled Channel method}
We apply the complex $G$-matrix interaction CEG07 to analyze the channel-coupling effect on elastic scattering and the energy dependence of the inelastic cross section through the MCC calculations.

\begin{widetext}
The coupled-channel (CC) equations for the radial component of the wave functions, $\chi^{(J)}_{\alpha L}(R)$ for a given total angular momentum of the projectile-target scattering system $J$, are written as, 
\begin{equation}
\left[
-\frac{\hbar^2}{2\mu} \frac{d^2}{dR^2}
+\frac{\hbar^2 L(L+1)}{2\mu R^2}
- E_{\alpha} 
\right] \chi^{(J)}_{\alpha L}(R) 
=-\sum_{\beta, L'}{F^{(J)}_{\alpha L, \beta L'}(R) \chi^{(J)}_{\beta L'}(R)},
\label{eq:cc}
\end{equation} 
where, $\mu$ is the reduced mass of the scattering system.
The suffix $\alpha$ for the radial wave function $\chi^{(J)}_{\alpha L}(R)$ denotes the channel number designated by the intrinsic spins of colliding two nuclei $I_1$ and $I_2$, the channel spin $S$ defined by the vector coupling of $I_1$ and $I_2$, and the sum of the excitation energies of the two nuclei $\epsilon_\alpha= \epsilon_1 + \epsilon_2$.
Namely, $\chi^{(J)}_{\alpha L}(R)\equiv \chi^{(J)}_{\alpha S(I_1 I_2) L}(R)$ explicitly.
Here, we assign $\alpha = 0$ to the entrance (elastic) channel. 
$E_{\alpha} = E_{\rm{c.m.}}-\epsilon_\alpha$ is the center-of-mass (c.m.) energy of the projectile-target relative motion in the channel $\alpha$, where $E_{\rm{c.m.}}$ is the c.m. energy in the elastic channel.
$L$ is the orbital angular momentum for the relative motion between the two nuclei which takes the values of $|J-S|\leq L \leq J+S$ for given $S$ and $J$.
Thus, the scattering channel is defined by a set of $\alpha$ and $L$ for a given $J$.
$F^{(J)}_{\alpha L, \beta L'}(R)$ represents the diagonal ($\alpha=\beta$ and $L=L'$) or coupling ($\alpha \neq \beta$ and/or $L \neq L'$) potential 
that is defined more explicitly~\cite{SYKT88,KAT02} by 
\begin{eqnarray}
F^{(J)}_{\alpha L, \beta L'}(R) \equiv F^{(J)}_{\alpha S(I_1 I_2) L,\beta S'(I'_1 I'_2) L'}(R) 
= \sum_{\lambda} i^{L + L' - \lambda} (-1)^{S + L' - J - \lambda} \hat{L} \hat{L'}\,
W(S L S' L': J \lambda) ( L 0 L' 0 | \lambda 0 ) \nonumber \\
\times 2 N_{I_1 I_2} N_{I'_1 I'_2} \left[ U^{(\lambda)}_{\alpha S(I_1 I_2), \beta S'(I'_1 I'_2)}(R) 
+ (-1)^{S} U^{(\lambda)}_{\alpha S(I_2 I_1), \beta S'(I'_1 I'_2)}(R) \right],
\label{eq:coupling}
\end{eqnarray}
where $N_{I_1 I_2} = [2(1 + \delta_{I_1I_2}\delta_{\epsilon_{1}\epsilon_{2}})]^{-\frac{1}{2}}$ and $\hat{L}=(2L+1)^{\frac{1}{2}}$.
$W(S L S' L: J \lambda)$ and $( L 0 L' 0 | \lambda 0 )$ denote the Racah and Clebsch-Gordan coefficients, respectively.
The second term on the right hand side appears for scattering of identical nuclei as in the present case of the $^{12}$C + $^{12}$C 
system~\cite{ITO01,KAT02}.

In Eq.~(\ref{eq:coupling}), $U^{(\lambda)}_{\alpha S(I_1 I_2), \beta S'(I'_1 I'_2)}(R)$ is the intrinsic component of the diagonal or coupling potential with the multipolarity of rank $\lambda$, that only contains nuclear structure information in channels $\alpha$ and $\beta$ and is irrelevant to the angular momenta $L$ and $J$ associated with the projectile-target relative motion.
It consists of the Coulomb and nuclear parts,
\begin{equation}
U^{(\lambda)}_{\alpha S(I_1 I_2), \beta S'(I'_1 I'_2)}(R)
= V^{(\lambda,  \rm{Coul.})}_{\alpha S(I_1 I_2), \beta S'(I'_1 I'_2)}(R) + U^{(\lambda,  \rm{Nucl.})}_{\alpha S(I_1 I_2), \beta S'(I'_1 I'_2)}(R)
\end{equation}
and they are obtained by the double folding of the Coulomb and nuclear parts of the nucleon-nucleon ($NN$) interaction, respectively, as
\begin{eqnarray}
V^{(\lambda, \rm{Coul.})}_{\alpha S(I_1 I_2), \beta S'(I'_1 I'_2)}(R) &=& \sqrt{4 \pi} \hat{S} \hat{S'} \hat{I_1} \hat{I_2}
\sum_{\lambda_1 \lambda_2}
\begin{Bmatrix}
I_1&I_2&S \\
I'_1&I'_2&S' \\
\lambda_1 &\lambda_2 &\lambda
\end{Bmatrix} \nonumber \\
&&\times \int{ \rho^{(\lambda_1, p)}_{I_1 I'_1}(r_1) 
               \rho^{(\lambda_2, p)}_{I_2 I'_2}(r_2)
               v^{(\rm{Coul.})}_{NN}(s)
\Big[ [       \mathscr{Y}_{\lambda_1}(\bm{\hat{r}}_1) 
       \otimes \mathscr{Y}_{\lambda_2}(\bm{\hat{r}}_2)]_{\lambda} 
       \otimes \mathscr{Y}_{\lambda}(\bm{\hat{R}}) \Big]_{00}
               d\bm{\hat{R}}d\bm{r}_1\bm{r}_2}, 
\label{eq:coulfold} \\
U^{(\lambda, \rm{Nucl.})}_{\alpha S(I_1 I_2), \beta S'(I'_1 I'_2)}(R) &=& \sqrt{4 \pi} \hat{S} \hat{S'} \hat{I_1} \hat{I_2}
\sum_{\lambda_1 \lambda_2}
\begin{Bmatrix}
I_1 &I_2 &S \\
I'_1&I'_2&S' \\
\lambda_1 &\lambda_2 &\lambda
\end{Bmatrix} \nonumber \\
&&\times \bigg \{ \int{ \rho^{(\lambda_1)}_{I_1 I'_1}(r_1) 
                  \rho^{(\lambda_2)}_{I_2 I'_2}(r_2)
                  v^{(\rm{D})}_{NN}(s, \rho, \epsilon )
\Big[ [\mathscr{Y}_{\lambda_1}(\bm{\hat{r}}_1) \otimes \mathscr{Y}_{\lambda_2}(\bm{\hat{r}}_2)]_{\lambda} \otimes \mathscr{Y}_{\lambda}(\bm{\hat{R}}) \Big]_{00} d\bm{\hat{R}} d\bm{r}_1 d\bm{r}_2} \nonumber \\
&& + \int{ \hat{j}_1(k^{\rm{eff}}_F(p)s) \rho^{(\lambda_1 )}_{I_1 I'_1}(p)
           \hat{j}_1(k^{\rm{eff}}_F(t)s) \rho^{(\lambda_2 )}_{I_2 I'_2}(t) 
           v^{(\rm{EX})}_{NN}(s, \rho, \epsilon ) 
           } \nonumber \\
&&\ \ \ \ \ \ \ \times \exp{\{ \frac{iM\bm{k}(R) \cdot \bm{s}}{\mu} \}}
               \Big[ [\mathscr{Y}_{\lambda_1}(\bm{\hat{p}}) 
               \otimes \mathscr{Y}_{\lambda_2}(\bm{\hat{t}})]_{\lambda} 
               \otimes \mathscr{Y}_{\lambda}(\bm{\hat{R}}) \Big]_{00} d\bm{\hat{R}} d\bm{p} d\bm{s}
\ \ \bigg \},
\label{eq:nuclfold}
\end{eqnarray}
\end{widetext}
where, $\bm{s} = \bm{R} - \bm{r_1} + \bm {r_2}$, $\bm{p} = \bm{r}_1 + \frac{1}{2}\bm{s}$, $\bm{t} = \bm{r}_2 - \frac{1}{2}\bm{s}$, and $\mathscr{Y}_{LM}(\hat{\bm{r}}) = i^{L}Y_{LM}(\hat{\bm{r}})$. 
In this expression, the Wigner 9-$j$ symbol is introduced.
$\epsilon = E/A$ is the incident energy per nucleon and $M$ is the nucleon mass. 
Here, the quantities $\rho^{(\lambda)}_{I I'}(r) $ represents the $\lambda$-rank multipole component of the diagonal or transition density of the projectile or target nucleus that is defined as 
\begin{equation}
\rho_{I m, I' m'}(\bm{r}) = 
\sqrt{4 \pi} \sum_{\lambda \nu} (I' m' \lambda \nu | I m) \rho^{(\lambda)}_{I I'}(r)\mathscr{Y}^{*}_{\lambda \nu}(\bm{\hat{r}}),
\label{eq:density}
\end{equation}
where $m$ and $m'$ are the z-component of $I$ and $I'$, respectively.
Note that $\rho^{(\lambda, p)}_{I I'}(r)$ with the superscript ($p$) in Eq.~(\ref{eq:coulfold}) represents the proton part of the density to be used in the Coulomb part of the folding potential.  
$v^{(\rm{Coul.})}_{NN}$ is the $NN$ Coulomb interaction, whereas $v^{(D)}_{NN}$ and $v^{(EX)}_{NN}$ are the direct and exchange parts of the nuclear interaction, respectively,  for which we adopt the complex $G$-matrix interaction CEG07 and they are written as 
\begin{equation}
v_{\rm{D, EX}}=\pm \frac{1}{16}v^{00}+\frac{3}{16}v^{01}+\frac{3}{16}v^{10} \pm \frac{9}{16}v^{11},
\label{eq:vst}
\end{equation}
in terms of the spin-isospin components $v^{ST}$ ($S$ = 0 or 1 and $T$ = 0 or 1) of the CEG07 interaction. 

The effective $NN$ interaction actually used in the preset MCC calculation is the CEG07b interactions~\cite{CEG07,FUR09}. 
The CEG07b includes the three-body force (TBF) effect that is found to be essentially important to predict proper shape and strength of the nucleus-nucleus interaction that are consistent with the observed elastic scattering data~\cite{FUR09R,FUR09,FUR10}. 

In the exchange part of Eq.~(\ref{eq:nuclfold}), $k(R)$ is the local momentum of the nucleus-nucleus relative motion defined by 
\begin{equation}
k^2(R)=\frac{2\mu}{\hbar^2}[E_{\rm{c.m.}}-{\rm{Re}}U^{(0, \rm{Nucl.})}_{0, 0}(R)
-V^{(0, \rm{Coul.})}_{0, 0}(R)], 
\label{eq:kkk}
\end{equation}
and the exchange part of the diagonal and coupling potentials
is calculated self-consistently on the basis of the local energy approximation through Eq.~(\ref{eq:kkk}).
Here, the local momentum is evaluated with the use of the nuclear and Coulomb potentials, Re~$U^{(0, \rm{Nucl.})}_{0, 0}$ and $V^{(0, \rm{Coul.})}_{0, 0}$, in the elastic channels, because the incident energies considered in the present paper is so high that the difference of the potentials between the elastic and inelastic channels is negligible in evaluating the local momentum.  
Note that $U^{(0, \rm{Nucl.})}_{0, 0}$ and $V^{(0, \rm{Coul.})}_{0, 0}$ in Eq.~(\ref{eq:kkk}) are the abbreviations of the $\lambda$ = 0 component of the nuclear and Coulomb potentials for the elastic channel defined by Eqs.~(\ref{eq:nuclfold}) and (\ref{eq:coulfold}), respectively.
In Eq.~(\ref{eq:nuclfold}), $\hat{j}_1(k^{\rm{eff}}_F(x)s) \equiv \frac{3}{k^{\rm{eff}}_F(x)s}j_1(k^{\rm{eff}}_F(x)s)$, where $k^{\rm{eff}}_{F}$ is the effective Fermi momentum \cite{CAM78} defined by
\begin{equation}
k^{\rm{eff}}_{F} 
=\Big( (3\pi^2 \rho )^{2/3}+\frac{5C_{\rm{s}}[\nabla\rho]^2}{3\rho^2}
+\frac{5\nabla ^2\rho}{36\rho} \Big)^{1/2}, \;\; 
\label{eq:kf}
\end{equation}
where we adopt $C_{\rm{s}} = 1/4$ following Ref.~\cite{KHO01}. 
The exponential function in Eq.~(\ref{eq:nuclfold}) is approximated by the leading term of the multipole expansion, namely the spherical Bessel function of rank 0, $j_{0} (\frac{Mk(R)s}{\mu})$,  following the standard prescription~\cite{BRI77,ROO77,BRI78,CEG83,CEG07,MIN10}.

The present $G$-matrix interaction, CEG07, depends on the density of nuclear medium and we have to specify the density to be used in the above folding-model calculations.
We employ the so-called frozen-density approximation (FDA)~\cite{FUR09} for evaluating the local density $\rho$ in Eq.~(\ref{eq:nuclfold}).
In the FDA, the density-dependent $NN$ interaction is assumed to feel the local density defined as the sum of the densities of the projectile and target nuclei; 
\begin{eqnarray}
\rho &=& \rho^{(\rm P)}+\rho^{(\rm T)}
\end{eqnarray}
In calculating the diagonal  ($I_1=I'_1$, $I_2=I'_2$) potential,
$U^{(\lambda, \rm{Nucl.})}_{\alpha S(I_1 I_2), \alpha S(I_1 I_2)}(R)$, we use the monopole ($\lambda=0$) component of the nucleon density defined by Eq.~(\ref{eq:density}) in the corresponding states of the projectile and the target nuclei, 
\begin{eqnarray}
\rho^{(\rm P)} &=& \rho^{(0)}_{I_1 I_1} \; , \;\;\;\;\; \rho^{(\rm T)} = \rho^{(0)}_{I_2 I_2} \; ,
\end{eqnarray}
that are nothing but the normal nucleon density in the states,
while in calculating the coupling potential $U^{(\lambda, \rm{Nucl.})}_{\alpha S(I_1 I_2), \beta S'(I'_1 I'_2)}(R)$, we use the
average of the nucleon densities in the initial and final states for each nucleus~\cite{ITO01,KAT02};
\begin{eqnarray}
\rho^{(\rm P)} &=& \frac12 \left\{ \rho^{(0)}_{I_1 I_1}+ \rho^{(0)}_{I'_1 I'_1} \right\} \; , \\
\rho^{(\rm T)} &=& \frac12 \left\{ \rho^{(0)}_{I_2 I_2}+ \rho^{(0)}_{I'_2 I'_2} \right\} \;.
\end{eqnarray}
The local densities are evaluated at the position of each nucleon for the direct part and at the middle point of the interacting nucleon pair for the exchange part following the preceding works~\cite{KHO00,KAT02}. 
The FDA has widely been used also in the standard DFM calculations~\cite{KHO94, KHO97, KAT02, SAT79, FUR09} 
and it was proved that the FDA was the most appropriate prescription for evaluating the local density
in the DFM calculations with realistic complex $G$-matrix interactions~\cite{FUR09}. 

The imaginary part of the calculated potential is multiplied by a renormalization factor $N_W$, the value of which is
the only free parameter in the present folding model.
In the previous analyses~\cite{FUR09R, FUR09}, its values were determined so as to reproduce the experimental data on the elastic-scattering cross sections to be compared with the calculated ones. 
However, there exist no experimental data to be compared with the calculations in the high energy region $E/A =$ 100 -- 400 MeV and we fix the $N_W$ value to unity unless otherwise mentioned as in Ref.~\cite{FUR10}.

We discuss the channel-coupling effect not only in the calculated cross sections but also in terms of the dynamical polarization potential (DPP)~\cite{DNR}.
The DPP in the elastic channel ($\alpha$ = 0) discussed in the present paper is the so-called {\it wave-function equivalent DPP}~\cite{SAK83,SAK86,SAK87} that is defined by
\begin{equation}
\Delta U^{(J)}_{\rm{DPP}}(R) = \sum_{\beta\neq 0, L'}{F^{(J)}_{0 J, \beta L'}(R) \chi^{(J)}_{\beta L'}(R)/\chi^{(J)}_{0 J}(R)}\; , 
\label{eq:dpp}
\end{equation}
where we use the fact that $I_1$ = $I_2$ = 0 and $L=J$ for the elastic channel in the present $^{12}$C~+~$^{12}$C scattering.

\section{Result and Discussion}

\subsection{Channel coupling effect on the elastic scattering}
We apply the MCC method with the CEG07 $G$-matrix interaction to the $^{12}$C + $^{12}$C elastic and inelastic scatterings at four incident energies per nucleon, $E/A =$ 100, 200, 300, and 400 MeV and first analyze the energy dependence of the channel coupling effect on the elastic scattering.
In the present MCC calculations, the single and mutual excitations of $^{12}$C to the $2_{1}^{+}$ (4.44 MeV), $0_{2}^{+}$ (7.65 MeV), $3_{1}^{-}$ (9.64 MeV), 
and $2_{2}^{+}$ (10.3 MeV)~\footnote{
The excitation energy, 10.3 MeV, of the $2_{2}^{+}$ state adopted here is slightly higher than that of the latest publication, 9.84 $\pm$ 0.06 MeV~\cite{ITO11}. However, the difference is completely negligible in the high-energy scattering studied in the present paper.
}
 excited states are taken into account. 
The diagonal and transition densities among the ground state and those excited states are taken from Ref.~\cite{KAM81} that were obtained by the 3$\alpha$-RGM (Resonance Group Method) calculation~\cite{FUJ80}. 
In this paper, we call the CC calculation that takes account of the full combination of excited states of the projectile and target nuclei as the full-CC calculation. 
However, the single excitation to the $2_{1}^{+}$ state are found to play a dominant role in the elastic and inelastic scattering discussed here. 

\begin{figure}[th]
\begin{center}
\includegraphics[width=6cm]{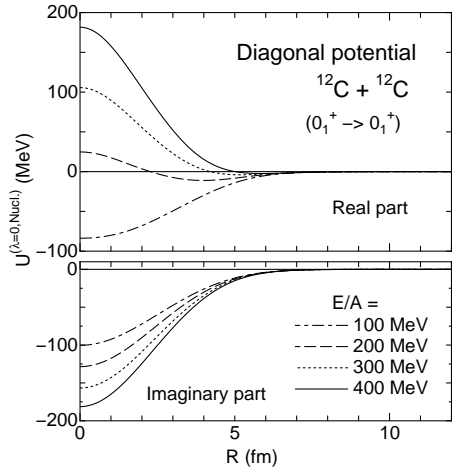}
\caption{\label{fig:01} The real and imaginary part of the folding potential in the elastic channel of the $^{12}$C + $^{12}$C system calculated at $E/A =$ 100, 200, 300, and 400 MeV.}
\end{center} 
\end{figure}

\begin{figure}[th]
\begin{center}
\includegraphics[width=6cm]{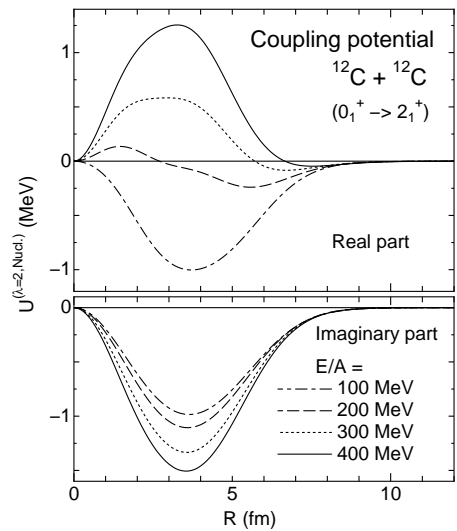}
\caption{\label{fig:02} The real and imaginary part the coupling potential (nuclear part defined in Eq.~(\ref{eq:nuclfold}) ) between the elastic channel and the $2_{1}^{+}$ single-excitation channel of the $^{12}$C + $^{12}$C system calculated at $E/A =$ 100, 200, 300, and 400 MeV.}
\end{center} 
\end{figure}

First, we plot the energy dependence of the real and imaginary parts of the 
diagonal potential in Fig.~\ref{fig:01}~\footnote{
A minor difference of the potentials shown in Fig.~\ref{fig:01} in the present paper and those given in Fig.~3 of Ref.~\cite{FUR10} is due to the difference in the adopted density for $^{12}$C and in the position where the local density is evaluated.
}. 
In the energy evolution, the real part of the folding potential in the elastic channel changes its sign between $E/A =$ 200 and 300 MeV, which was already reported in the previous work~\cite{FUR10} and referred to as the {\em the attractive to repulsive transition}.
The experimental confirmation of the transition predicted by the microscopic folding model will be a big challenge~\cite{GUO12}.
Figure~\ref{fig:02} shows the energy evolution of the coupling potential between the elastic channel and the $2_{1}^{+}$ single-excitation channel.
The real part of the coupling potential changes its sign between $E/A =$ 200 and 300 MeV in the same manner as in the case of the elastic-channel potential, whereas the strength of the imaginary part slowly and monotonically increases with the increase of the energy that also follows the trend of the elastic-channel potential.
One should note here that the real and imaginary parts of the coupling potential have similar strength at $E/A =$ 100 MeV, while the real to imaginary ratio as well as their relative sign drastically changes with the increase of the energy, which will be a very important point in understanding the energy dependence of the DPP as will be discussed later.


\begin{figure}[tbh]
\begin{center}
\includegraphics[width=6.5cm]{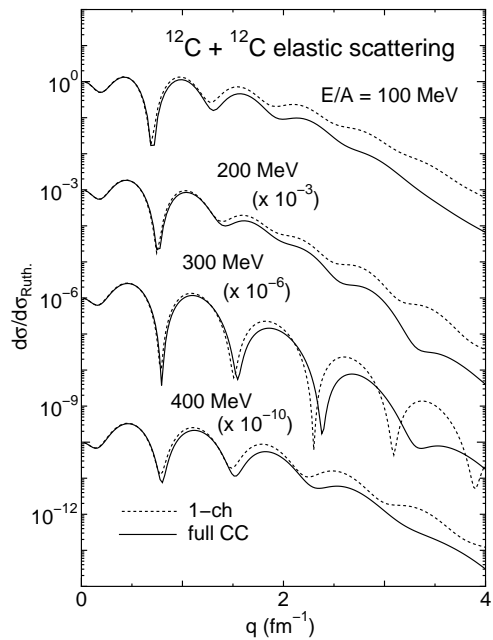}
\caption{\label{fig:03} The Rutherford ratio of the elastic-scattering cross sections displayed as the functions of the momentum transfer $q$.
The dotted and solid curves are the results of the single-channel and full-CC calculations, respectively. 
}
\end{center} 
\end{figure}
Figure~\ref{fig:03} shows the angular distributions of the $^{12}$C~+~$^{12}$C elastic cross sections calculated at the four incident energies.
The relativistic-kinematics correction has been made in all the calculations presented in this paper.
The dotted and solid curves are the results of the single-channel and full-CC calculations, respectively. 
A sizable channel coupling effect is clearly seen in the elastic cross sections at all incident energies including the highest energy.
It is found that the dominant contribution to the channel-coupling effects on the elastic scattering comes from the $2_{1}^{+}$ single-excitation channel.
In the comparison of the single-channel calculation with the full-CC one, one notices that the diffraction pattern of the cross sections slightly shifts backward and the cross sections decreases at large angles by the channel-coupling effect.
Although the effect on the cross sections looks similar to all the incident energies, the contents of the effect are very different from each other as will be discussed below in terms of the DPP.

\subsection{Dynamical polarization potential}
\begin{figure}[thb]
\begin{center}
\includegraphics[width=6cm]{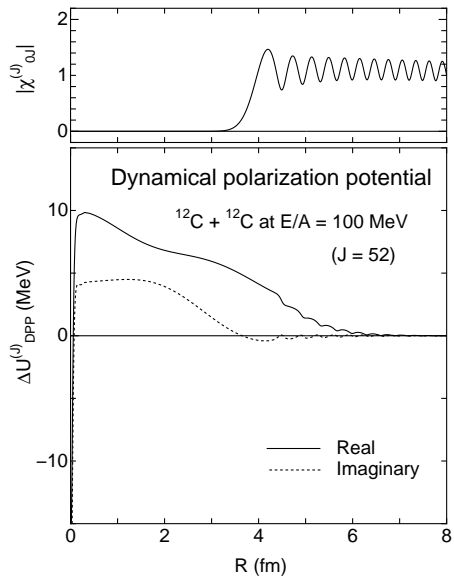}
\caption{\label{fig:04} The DPP for $J_{\rm gr}=$ 52 obtained by the full-CC calculation at $E/A =$ 100 MeV, plotted along with the absolute value of the radial wave function in the elastic channel. The solid and dotted curves are the real and imaginary parts of the DPP, respectively. 
}
\end{center} 
\end{figure}
\begin{figure}[thb]
\begin{center}
\includegraphics[width=6cm]{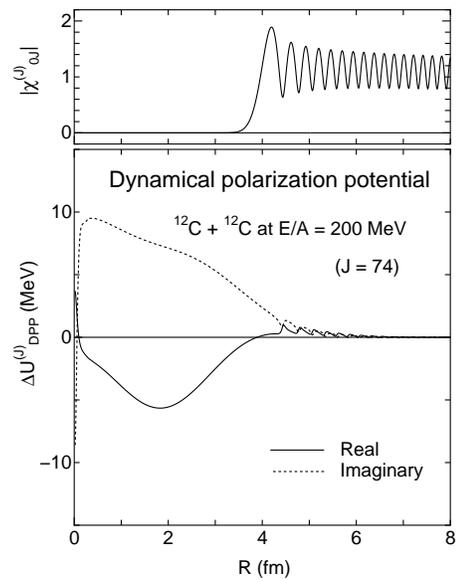}
\caption{\label{fig:05} Same as Fig.~\ref{fig:04} but for $J_{\rm gr}=$ 74 at $E/A =$ 200 MeV.}
\end{center} 
\end{figure}
\begin{figure}[thb]
\begin{center}
\includegraphics[width=6cm]{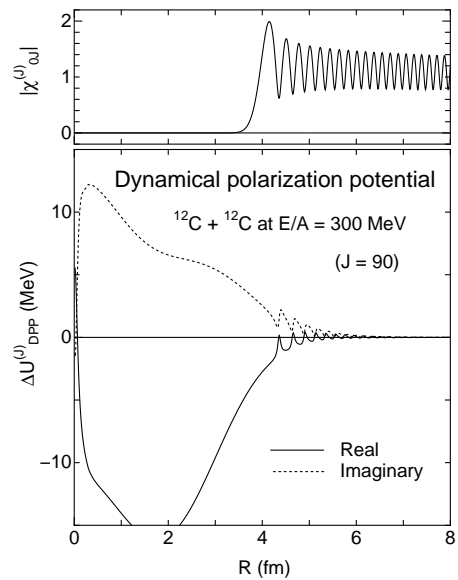}
\caption{\label{fig:06} Same as Fig.~\ref{fig:04} but for $J_{\rm gr}=$ 90 at $E/A =$ 300 MeV.}
\end{center} 
\end{figure}
\begin{figure}[thb]
\begin{center}
\includegraphics[width=6cm]{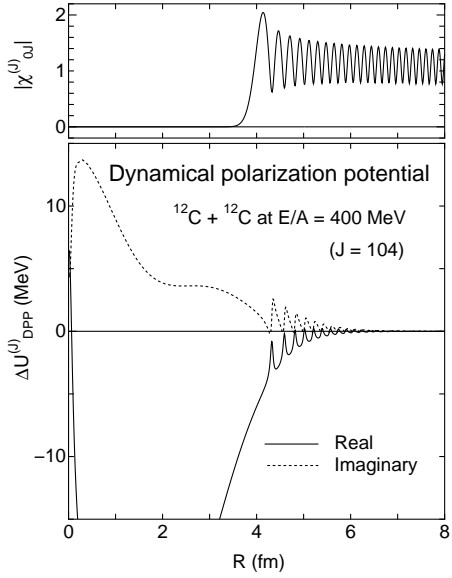}
\caption{\label{fig:07} Same as Fig.~\ref{fig:04} but for $J_{\rm gr}=$ 104 at $E/A =$ 400 MeV.}
\end{center} 
\end{figure}
Next, we investigate the channel-coupling effect on the elastic scattering in terms of the dynamical polarization potential (DPP) defined by Eq.~(\ref{eq:dpp}).
By definition, the DPP is $J$ dependent and we calculate the DPP for a $J$ value close to the grazing $J$ value defined by 
\begin{equation}
J_{\rm gr}=b_{\rm gr}\sqrt{2 \mu E_{\rm c.m.}}/\hbar\; . 
\end{equation}
Here, $b_{\rm gr}$ denotes the grazing impact parameter and we take $b_{\rm gr} \simeq  4$ fm in the present $^{12}$C~+~$^{12}$C system. 
Figures~\ref{fig:04} to \ref{fig:07} show the real and imaginary parts of the DPP for $E/A =$ 100, 200, 300 and 400 MeV calculated at $J_{\rm gr}$ for each incident energy, along with the absolute value of the elastic-channel wave function, $\big| \chi^{(J)}_{0 J}(R) \big|$ . 
For the grazing $J$ values, the elastic-channel wave function has visible magnitudes outside $R\approx 3.5$ fm and reaches its maximum around the grazing impact parameter positions $R \approx 4$ fm, as seen in the figures.
The $J$ dependence of the DPP will be discussed later.

The oscillation of the DPP seen in the region outside the grazing distances is mainly due to the oscillation of the elastic-channel wave function that appears in the denominator in the definition of the DPP (Eq.~(\ref{eq:dpp})).
The oscillation, however, does not give rise to any anomaly in the calculated cross sections because the DPP multiplied by the elastic-channel wave function in the CC equations is a smooth function of the radial variable $R$.
It should be noted that the DPP at short distances, say less than about 3 fm, plays little role in the scattering because of the repulsion of the centrifugal barrier on one hand and, on the other hand, because of the strong absorption in the internal region~\cite{FUR10}, that make the magnitude of the elastic-channel wave function negligibly small.
Thus, hereafter, we discuss the DPP only in the region of $R \geq 3$ fm outside the insensitive domain.

\begin{figure}[th]
\begin{center}
\includegraphics[width=6cm]{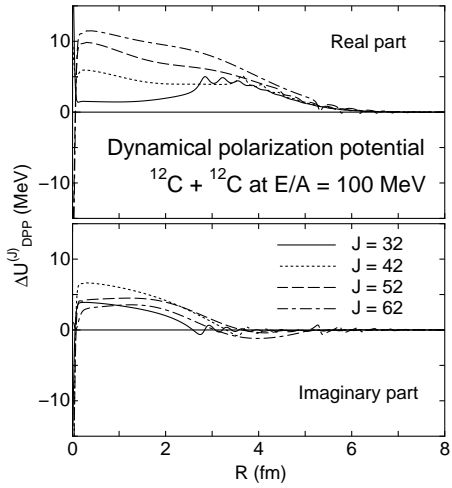}
\caption{\label{fig:08} Real (upper) and imaginary (lower) parts of the DPP at $E/A =$ 100 MeV.
The solid, dotted, dashed, and dot-dashed curves are the results at $J$ = 32, 42, 52, and 62, respectively. 
}
\end{center} 
\end{figure}
\begin{figure}[th]
\begin{center}
\includegraphics[width=6cm]{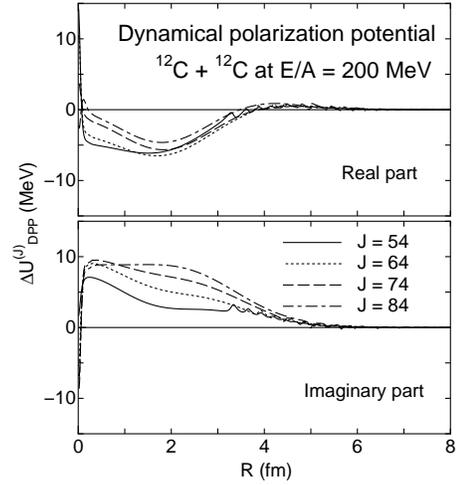}
\caption{\label{fig:09} Same as Fig.~\ref{fig:08} but at $E/A =$ 200 MeV and $J$ = 54, 64, 74, and 84.}
\end{center} 
\end{figure}
\begin{figure}[th]
\begin{center}
\includegraphics[width=6cm]{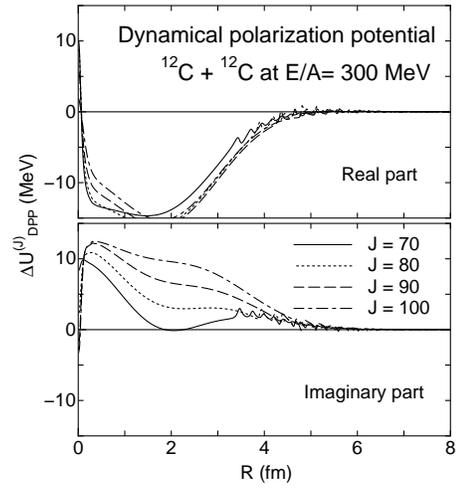}
\caption{\label{fig:10} Same as Fig.~\ref{fig:08} but at $E/A =$ 300 MeV and $J$ = 70, 80, 90, and 100.}
\end{center} 
\end{figure}
\begin{figure}[th]
\begin{center}
\includegraphics[width=6cm]{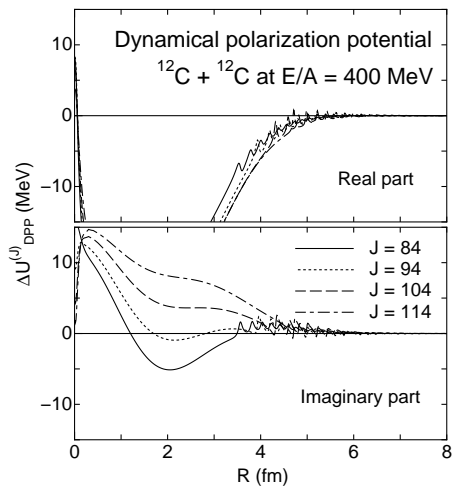}
\caption{\label{fig:11} Same as Fig.~\ref{fig:08} but at $E/A =$ 400 MeV and $J$ = 84, 94, 104, and 114.}
\end{center} 
\end{figure}
As the increase of the incident energy, the real part of the DPP shows the rapid transition from positive (repulsive) to negative (attractive) around $E/A =$ 200 MeV, whereas the imaginary part changes its sign from negative to positive in the surface region.
The characteristic energy dependence of the real and imaginary parts of the DPP calculated at the grazing $J$ values still persists to other $J$ values, as shown in Figs~\ref{fig:08} to \ref{fig:11} that show the DPP calculated for various $J$ values around the $J_{\rm gr}$ value for each incident energy.
One can clearly see that the $J$ dependence of the DPP is small outside the insensitive radial region $R\geq 3$ fm for all the incident energies and the characteristic trend of the energy dependence of DPP is almost independent of the $J$ value.

\subsection{Relation between complex coupling potential and DPP }
It is well known that the sign and the strength of the DPP have a close relation to the relative sign and the strength of the real and imaginary parts of the coupling potential~\cite{FRA81,SAK87,SAK86}.
Therefore, the characteristic energy evolution of the DPP can be easily understood through the energy dependence of the complex coupling potential shown in Fig,~\ref{fig:02}.
For example, the negative (attractive) sign of the real part of DPP at $E/A =$ 300 and 400 MeV is the result of the different sign of the real and imaginary parts of the coupling potential at those energies, whereas the positive sign of the imaginary part of DPP at $E/A =$ 200 and 300 MeV can be understood by the dominance of the imaginary coupling at the sensitive region ($R\geq 3$ fm) at these energies~\cite{SAK87}.

To understand the close relation between the signs of the complex DPP and those of the complex coupling potentials, we make use of the following  simplified model adopted in Refs.~\cite{SAK86,SAK87} in the analysis.
In this model, one assumes the same radial form factor to the real and imaginary parts of the coupling potential in order to clarify the close relation between the complex coupling potential and the complex DPP and discusses the DPP evaluated in the second-order approximation of the reaction processes. 
Here, we define the complex coupling potential between the elastic channel ($\alpha=0$) and a channel $\beta$ having the same radial form factor which we write symbolically as
\begin{eqnarray}
U_{0 \beta}(R) &=& (N_{\rm{V}}+iN_{\rm{W}})f_{0 \beta}(R) \; .
\label{eq:CP}
\end{eqnarray}
$f_{0 \beta}(R)$ denotes the radial form factor of the coupling potential that is taken to be a real function of $R$, whereas $N_{\rm{V}}$ and $N_{\rm{W}}$ represent the strength parameters for the real and imaginary parts of the complex coupling potential, respectively.

Then, the DPP generated by the complex coupling between the elastic channel and various excited channels \{$\beta$\} in the second-order perturbation
theory will be written symbolically as
\begin{eqnarray}
\Delta U_{\rm DPP} &=& \sum_{\beta} U_{0 \beta} \hat G^{(+)}_\beta U_{\beta 0} \\
                   &=& (N_{\rm{V}}+iN_{\rm{W}})^2 \sum_{\beta} f_{0 \beta} \hat G_\beta f_{\beta 0} \\
                   &\equiv & (N_{\rm{V}}+iN_{\rm{W}})^2 (\Delta v + i\Delta w) \label{eq:DPP} \; ,
\end{eqnarray}
where $\hat G^{(+)}_\beta$ represents the Green function operator (propagator) in the $\beta$ channel and $\Delta v + i\Delta w$ is the complex DPP generated by the real coupling potential $f_{0 \beta}(R)$ with a unit strength.
It is well known~\cite{DNR,LTS77,KUB81} that the DPP generated by the real coupling potential having a moderate strength corresponding to the normal collective excitations is dominated by the imaginary part of an absorptive nature ($\Delta w < 0$) and the real part of the DPP is rather small or even negligible, $|\Delta w| \gg |\Delta v|\approx 0$.
In this situation, the real and imaginary parts of the DPP given by Eq.~(\ref{eq:DPP}), $\Delta V_{\rm{DPP}}$ and $\Delta W_{\rm{DPP}}$, generated by the complex coupling $U_{0 \beta}(R)$  will be
\begin{eqnarray}
\Delta V_{\rm{DPP}}&\simeq  &-2N_{\rm{V}}N_{\rm{W}}\Delta w, \label{eq:dppreal} \\
\Delta W_{\rm{DPP}}&\simeq  &(N_{\rm{V}}^2-N_{\rm{W}}^2)\Delta w, \label{eq:dppimag}
\end{eqnarray}
in good approximation.\footnote{A more detailed discussion in the situation with $\Delta v \neq 0$ can be found in Refs.~\cite{SAK86,SAK87}}

The relations (\ref{eq:dppreal}) and (\ref{eq:dppimag}) clearly show the close relations between the real and imaginary parts of the DPP and the strength and sign of the real and imaginary coupling potential.
For instance, one can see that the existence of the imaginary coupling ($N_{\rm{W}} \neq 0$) is essential to generate the real part of DPP and that the same sign of the real and imaginary coupling potentials gives rise to a repulsive DPP ($\Delta V_{\rm{DPP}} > 0$) because of the fact that $\Delta w < 0$, whereas the similar strength of the real and imaginary coupling potentials leads to the pure real DPP ($\Delta W_{\rm{DPP}}\approx 0$), that was really the case of the repulsive DPP generated by the coupling to the breakup processes for the loosely-bound $^{6,7}$Li and $^{9}$Be projectile nuclei~\cite{SAK83,SAK86,SAK87}.

Based on the relations (\ref{eq:dppreal}) and (\ref{eq:dppimag}), one can clearly and easily understand the peculiar energy dependence of the real and imaginary parts of the DPP that we saw in Figs.~\ref{fig:04} to \ref{fig:07} in terms of the characteristic energy dependence of the coupling potential shown in Fig.~\ref{fig:02}.
At $E/A =$ 100 MeV, the similar strengths of the real and imaginary part of the coupling potential with the same negative sign ($N_{\rm{V}}\approx N_{\rm{W}} < 0$) lead to the strongly repulsive $\Delta V_{\rm{DPP}} > 0$ with a negligible imaginary part $\Delta W_{\rm{DPP}}\approx 0$ as seen in Fig.~\ref{fig:04}, whereas, at $E/A =$ 200 MeV, the weak real coupling ($N_{\rm{V}}\approx 0$) and the negative imaginary coupling still having a considerable strength ($N_{\rm{W}} < 0$) lead to the completely opposite situation to the $E/A =$ 100 MeV case, i.e. a negligible real DPP ($\Delta V_{\rm{DPP}}\approx 0$) with a sizable imaginary DPP having the positive sign ($\Delta W_{\rm{DPP}} > 0$) as we have seen in Fig.~\ref{fig:05}.
As one goes further to the higher energy of $E/A =$ 400 MeV, the real and imaginary coupling have similar strength but of opposite signs, which may correspond to the situation of $-N_{\rm{W}}\approx N_{\rm{V}} > 0$ in the present schematic model, which leads to the real DPP of the strongly attractive nature ($\Delta V_{\rm{DPP}} < 0$) with a relatively weak imaginary DPP ($\Delta W_{\rm{DPP}}\approx 0$), although the situation rather depends on the radial region as will be understood in the comparison of Figs.~\ref{fig:02} and \ref{fig:07}.

\subsection{Effect of DPP on the elastic cross sections}
Now, we make a comment about the channel coupling effect on the elastic-scattering cross sections shown in Fig.~\ref{fig:03}.
As was already mentioned concerning to the figure, the channel-coupling effects on the angular distribution of elastic scattering at four incident energies look very similar to each other but we will point out that the origin of the effect is very different from each other.

\begin{figure}[tbh]
\begin{center}
\includegraphics[width=8.5cm]{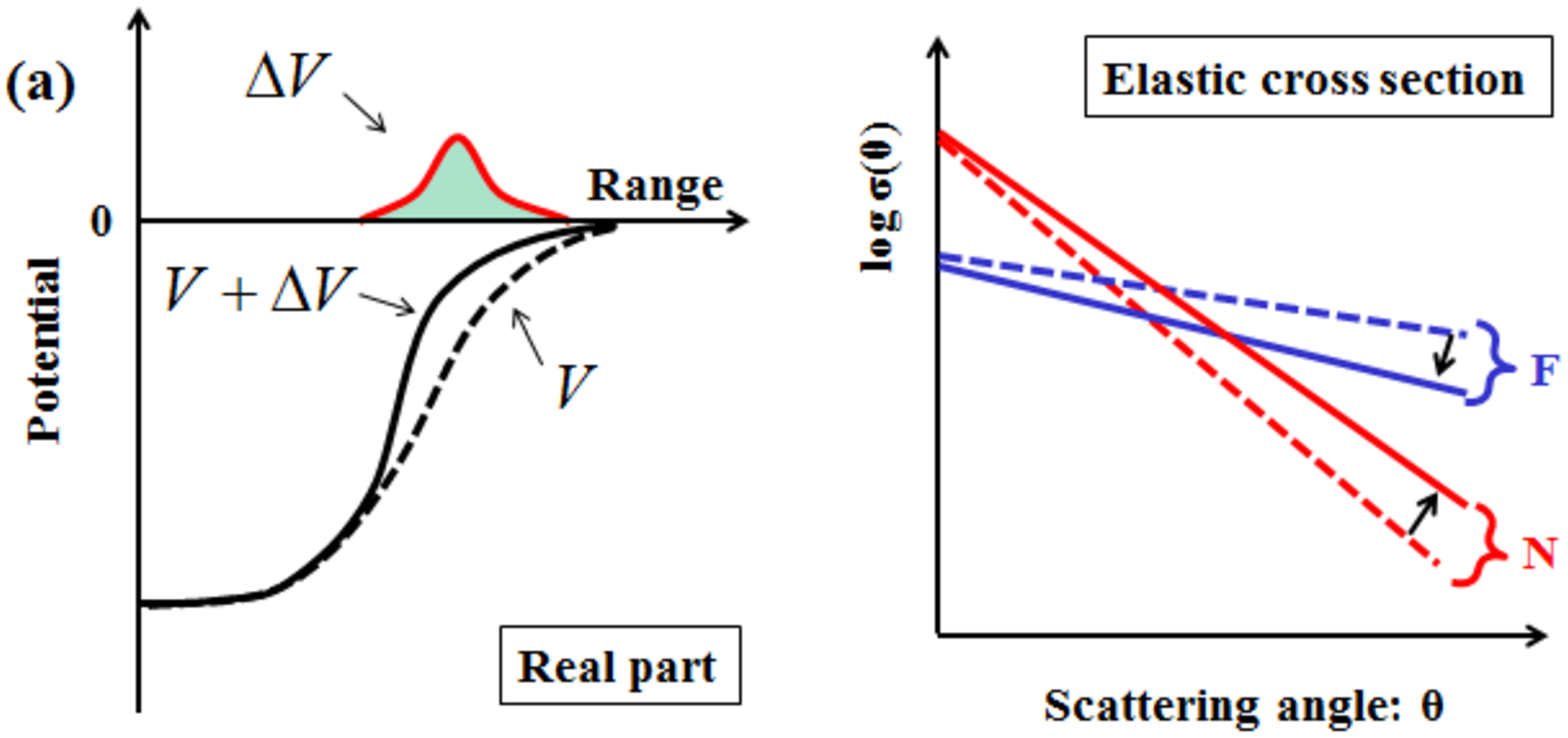}\\
\includegraphics[width=8.5cm]{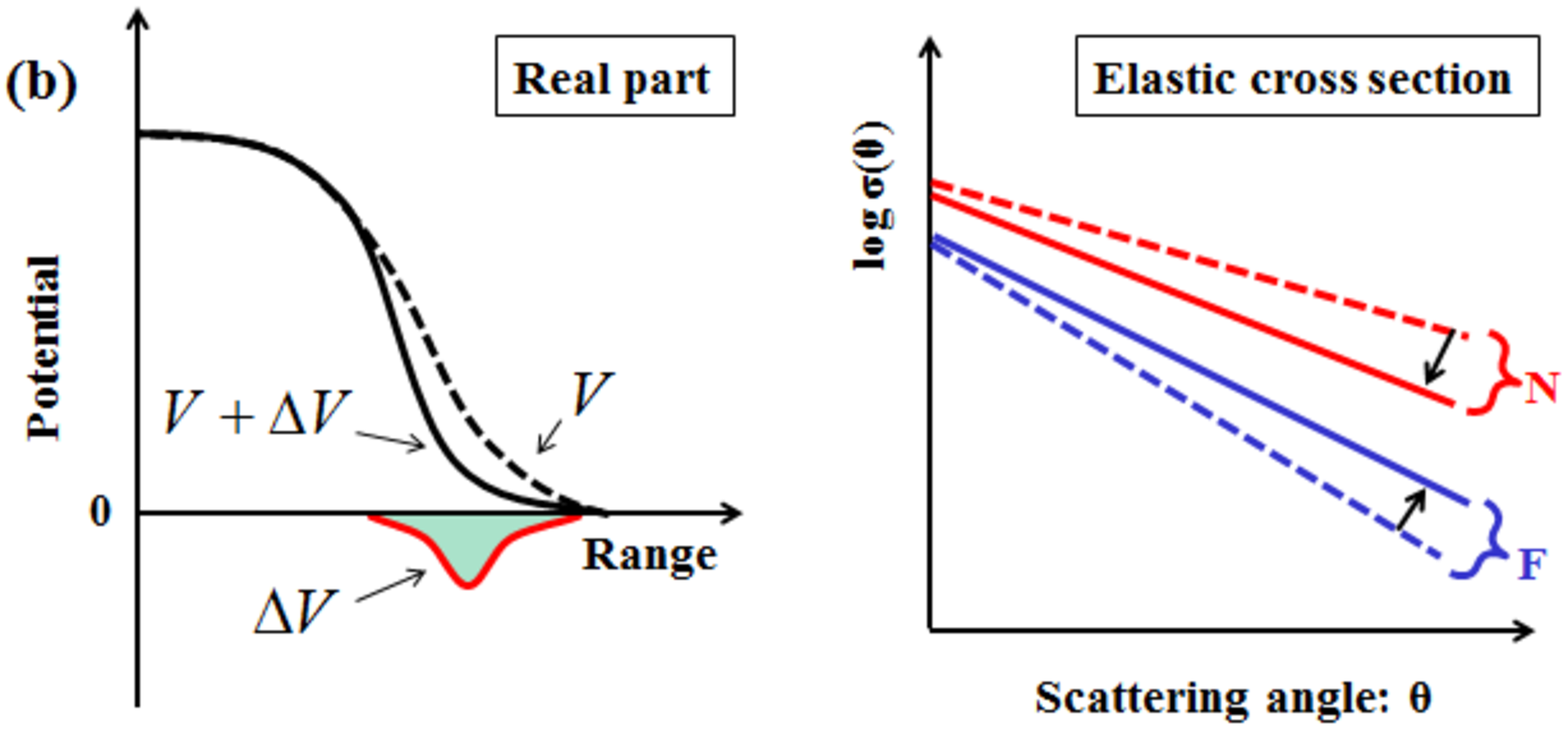}
\caption{\label{fig:12} (Color online) Schematic picture for the relation of the bare potential + DPP and the nearside and farside cross sections.
}
\end{center} 
\end{figure}
At the lowest energy $E/A =$ 100 MeV, the real part of the bare folding potential is strongly attractive potential that acts to swing the trajectory of the incoming projectile nucleus toward the opposite side of the target nucleus in the semiclassical picture.
This implies that, in terms of the nearside/farside decomposition picture~\cite{FUL75}, the attractive potential enhances the farside-scattering amplitude at backward angles that leads to a crossover around the middle angles with the nearside-scattering amplitude that dominates the scattering at forward angles as shown by the dashed lines in Fig.~\ref{fig:12}(a) (upper panel). 
At this energy, the real part of the DPP has positive sign (being of the repulsive nature) as we have seen in Fig.~\ref{fig:04}.
This implies that the sum of the attractive bare folding potential and the repulsive DPP ($V + \Delta V$; the solid curve in Fig.~\ref{fig:12}(a)) leads to a less attractive potential compared with the bare folding potential $V$ (the dashed curve).
This results in the hindrance of the farside amplitude and the enhancement of the nearside one (the red and blue solid lines in the right panel of Fig.~\ref{fig:12}(a)), which leads to the decrease of the coherent sum of the nearside and farside cross sections accompanied by the slightly backward shift of the diffraction, as seen by the change from the dotted curve to the solid one for the $E/A =$ 100 MeV case in Fig.~\ref{fig:03}.

The situation is completely opposite in the case of $E/A =$ 400 MeV, where the real part of the bare folding potential is strongly repulsive while the real part of the DPP has negative sign (being of the attractive nature) as we have seen in Fig.~\ref{fig:02} and \ref{fig:07}, which is schematically shown in the left panel of Fig.~\ref{fig:12}(b).
In such a situation, the elastic scattering is dominated by the nearside amplitude over the whole angular region because of the repulsive nature of the bare folding potential ($V > 0$) \cite{FUR10} but the addition of the attractive DPP ($\Delta V < 0$) lead to a less repulsive potential, as shown by the solid curve in the left panel of Fig.~\ref{fig:12}(b).
The nearside amplitude (red lines) generated by the less repulsive potential slightly decreases, while the farside amplitude (blue lines) slightly increases, which leads again to the decrease of the coherent sum of the two amplitudes.
This is what we have observed in Fig.~\ref{fig:03} in the case of $E/A =$ 400 MeV.

\subsection{Role of the real and imaginary coupling in the elastic and inelastic cross sections}
\begin{figure}[th]
\begin{center}
\includegraphics[width=6.5cm]{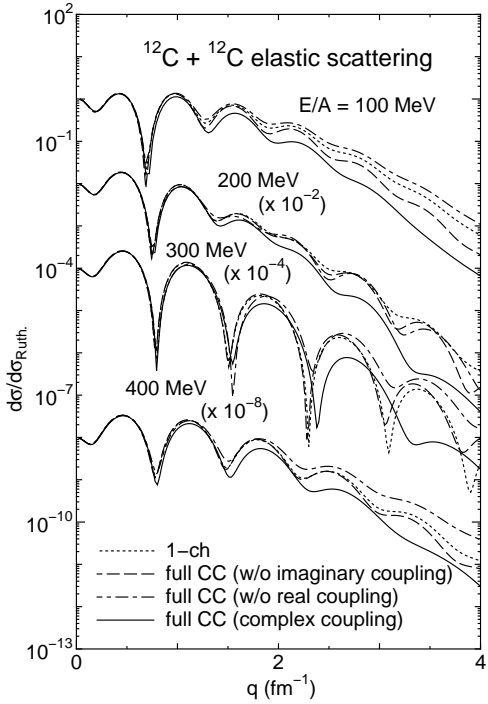}
\caption{\label{fig:13} The effects of the real and imaginary coupling potentials on the elastic cross section. 
The solid, dotted, dashed, and dot-dashed curves are the results by full CC with complex coupling, single channel, full CC without imaginary coupling, and full CC without real coupling calculations, respectively.
}
\end{center} 
\end{figure}
As discussed in the previous section, the basic idea of the relation between the channel-coupling effect on the elastic scattering and the strength/sign of the real and imaginary parts of the coupling potential can be understood in terms of the DPP through Eqs.(\ref{eq:dppreal}) and (\ref{eq:dppimag}) and we understand that both the real and imaginary parts of the coupling potential play important roles. 
Here, we investigate the individual roles of the real and imaginary parts of the coupling potential on the elastic and inelastic cross sections. 
To this end, we perform the CC calculations by switching off either the real part or the imaginary part of the coupling potential and see the effects on the elastic and inelastic cross sections.

Figure~\ref{fig:13} shows the calculated elastic cross sections for the $^{12}$C + $^{12}$C system at $E/A =$ 100 -- 400 MeV.
The dotted and solid curves show the results of the single-channel calculation and the full-CC ones with the complex coupling, respectively, 
which are the same as those in Fig.~\ref{fig:03}. 
The large difference between the dotted and solid curves shows the significant effects of the channel coupling with the complex coupling potentials.
The dot-dashed (dashed) curves show the results of the CC calculations without the real part (imaginary part) of the coupling potential.
In the cases without either the real or imaginary coupling potentials, the channel coupling effect becomes much smaller than the case with the complex coupling potentials, as seen in Fig.~\ref{fig:13}.
These results clearly indicate that both the real and imaginary parts of the complex coupling potential play important roles to give the large channel coupling effect on the elastic cross section. 

It is interesting to note that the elastic cross sections by the CC calculation without the real coupling potential (dot-dashed curves) are larger than those by the single-channel ones (dotted curves) at backward angles.
This can be understood through Eqs.~(\ref{eq:dppreal}) and (\ref{eq:dppimag}). 
Namely, the CC calculation with a pure imaginary coupling ($N_{\rm V}$ = 0, $N_{\rm W}\neq$ 0) leads to 
$\Delta V_{\rm DPP}\cong 0$ and $\Delta W_{\rm DPP}\cong - N^2_{\rm W}\, \Delta w >0$, 
which implies the simple decrease of the absorption with respect to the bare folding potential used in the single-channel calculation, $|W+\Delta W_{\rm DPP}| < |W|$, leading to the enhancement of cross sections at backward angles.

\begin{figure}[th]
\begin{center}
\includegraphics[width=6.5cm]{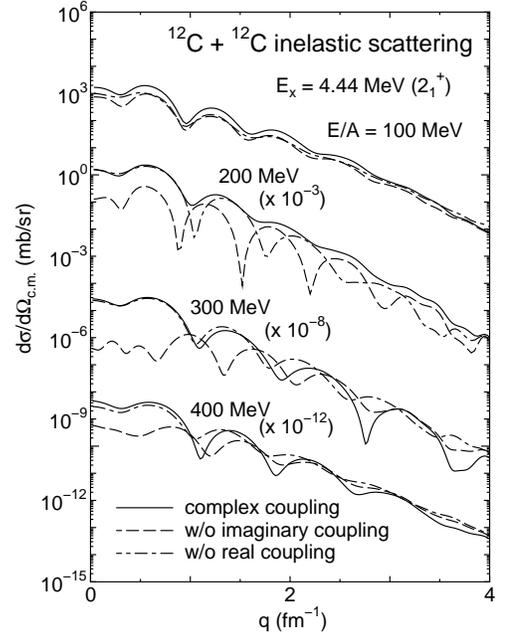}
\caption{\label{fig:14} The effects of the real and imaginary coupling potentials on the inelastic cross section. 
The solid, dashed, dot-dashed curves are the results by the full CC with complex coupling, full CC without imaginary coupling, and full CC without real coupling calculations, respectively.
}
\end{center} 
\end{figure}

Next, we show the results on the inelastic cross sections.
Figure~\ref{fig:14} shows the calculated cross sections of the $^{12}$C + $^{12}$C inelastic scattering to the $2_1^+$ single-excitation channel at the same incident energies, $E/A =$ 100 -- 400 MeV.
The solid curves show the results with the complex coupling potential, whereas the dot-dashed (dashed) curves show the results without the real part (imaginary part) of the coupling potential.
At $E/A =$ 100 MeV, the real and imaginary parts of the coupling potential have comparable contributions to the inelastic cross sections and their coherent sum (the solid curve for the complex coupling) has twice those individual contributions.

The situation completely changes as one goes to the higher energies, where the inelastic cross sections are dominated by the imaginary part of the coupling potential and the contribution of the real part of the coupling potential is quite small, particularly at $E/A =$ 300 MeV where the contribution is by two order of magnitude smaller than that of the imaginary part and almost negligible, as far as the cross sections at forward angles (i.e. for low momentum transfer (low-$q$) regions) is concerned.

The drastic energy dependence of the contribution of the real and imaginary coupling on the inelastic cross sections is found to have a close relation to the characteristic energy dependence of the complex coupling potential shown in Fig.~\ref{fig:02}.
To make the discussion clear and simple, we again make use of the simplified model, Eq.~(\ref{eq:CP}), for the coupling potential and assume that the inelastic transition occurs in the one-step process.
On this assumption, the inelastic cross section will be evaluated by the distorted-wave Born approximation (DWBA) and written symbolically as,
\begin{eqnarray}
\sigma_{\beta, 0} & = & 
        \Bigl| \bigl< \psi_\beta^{(-)} \bigl| (N_{\rm{V}}+iN_{\rm{W}})f_{\beta 0}(R) \bigr| \psi_0^{(+)} \bigr> \Bigr|^2 \nonumber \\
  & = &  ( N_{\rm{V}}^2 + N_{\rm{W}}^2 ) \Bigl| \bigl< \psi_\beta^{(-)} \bigl| f_{\beta 0}(R) \bigr| \psi_0^{(+)} \bigr> \Bigr|^2 \nonumber \\
  & \equiv &  \sigma_{\beta, 0}^{(\rm r)} + \sigma_{\beta, 0}^{(\rm i)}
\; ,
\label{eq:DWBA}
\end{eqnarray}
where $\psi_0^{(+)}$ and $\psi_\beta^{(-)}$ denote the distorted waves in the entrance and exit channels, respectively, and the first and second terms on the right hand side of Eq.~(\ref{eq:DWBA}) correspond to the contributions of the real and imaginary parts of the coupling potential to the inelastic cross section, respectively. 
By referring to Fig.~\ref{fig:02}, the real and imaginary parts of the coupling potential have comparable magnitude at $E/A =$ 100 MeV around the nuclear surface region, which may correspond to the situation as $|N_{\rm{V}}| \approx  |N_{\rm{W}}|$ in this model.
This leads to $ \sigma_{\beta, 0}^{(\rm r)} \approx \sigma_{\beta, 0}^{(\rm i)} \approx \frac12 \sigma_{\beta, 0}$ and this is just what we see in Fig.~\ref{fig:14} in the case of $E/A =$ 100 MeV.

\begin{figure}[th]
\begin{center}
\includegraphics[width=6.5cm]{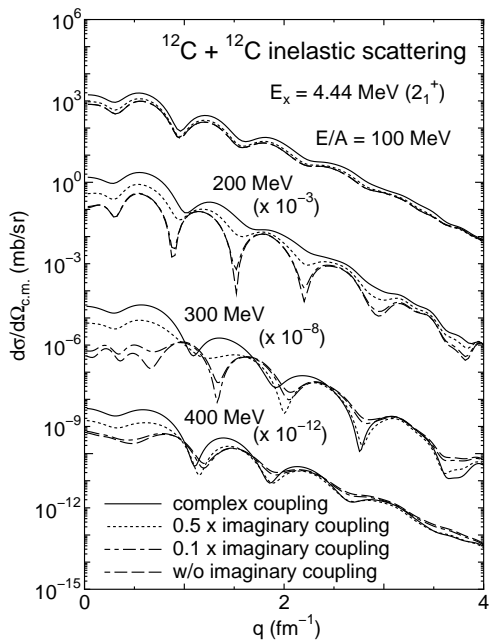}
\caption{\label{fig:15} The effect of the strength of the imaginary coupling potentials on the inelastic cross section. 
The solid, dotted, dot-dashed, and dashed curves are the results by the full CC with complex coupling, full CC with the imaginary coupling multiplied by 0.5, full CC with the imaginary coupling multiplied by 0.1, and full CC without imaginary coupling calculations, respectively.
}
\end{center} 
\end{figure}
As seen in Fig.~\ref{fig:02}, the magnitude of the real part of the coupling potential decreases as the increase of the incident energy and changes its sign around $E/A =$ 300 MeV and becomes positive at $E/A =$ 400 MeV except at the most periphery, while strength of the imaginary part monotonically increases with the increase of the energy.
These situations may correspond to the case that $|N_{\rm{V}}| \ll  |N_{\rm{W}}|$ in the present simplified model, which results in the negligible contribution from the real coupling to the inelastic cross section that is dominated by the imaginary coupling, 
$\sigma_{\beta, 0}^{(\rm r)} \ll \sigma_{\beta, 0}^{(\rm i)} \approx \sigma_{\beta, 0}$.
This precisely explains the situations what we observe in Fig.~\ref{fig:14} in the realistic CC calculations at $E/A =$ 200 -- 400 MeV as far as the cross sections at forward angles are concerned.

It should be note that the contributions of the real and imaginary coupling potential become comparable at large angles (at high-$q$) even at higher energies as seen in Fig.~\ref{fig:14}. 
This may also be understood within the present simplified model, because the cross section in high-$q$ region reflects the contribution of the coupling potentials at short distances where the real part has a strength comparable to that of the imaginary part, as shown in Fig.~\ref{fig:02}.

The imaginary part of the coupling potential has an important role for the inelastic cross section, especially at forward angles as shown in Fig.~\ref{fig:14}.
Therefore, we analyze the sensitivity of the strength of the imaginary coupling potential.
Figure~\ref{fig:15} shows the inelastic cross sections calculated with the artificial change of the strength of the imaginary coupling potential.
The dotted, dot-dashed and dashed curves in Fig.~\ref{fig:15} show the results with the imaginary coupling potential being multiplied by the factor of 0.5, 0.1 and 0.0 respectively, which are compared with the result with the original strength (the solid curves).
It is clearly seen that the inelastic cross section at forward angles is very sensitive to the strength of the imaginary part of the coupling potential and almost proportional to its square in the case of $E/A =$ 300 MeV where the contribution of the real coupling potential is negligible.
In other words, the measurement of the absolute magnitude of the inelastic cross sections at very forward angles around these incident energies will provide a crucial test for the validity of microscopic interaction models, including the present one based on the complex $G$-matrix CEG07, that predicts the shape and strength as well as their energy dependence of the complex coupling potential to be used in the reaction calculations. 
It is of particular importance to test the validity of the imaginary part of the coupling potential that should contain very complicated reaction processes via the so-called $Q$-space not included in the model space for reaction calculations under consideration.

\section{Conclusion}
The channel-coupling effect on the elastic and inelastic scattering of the $^{12}$C + $^{12}$C system is investigated with the microscopic coupled-channel (MCC) method using the complex $G$-matrix interaction CEG07 at $E/A =$ 100, 200, 300, and 400 MeV. 
The large effects of low-lying excitations are clearly seen in all the incident energies investigated, despite the high incident energies.
The present MCC method predicts the drastic energy dependence of the shape and strength of the complex coupling potential, that is very similar to the energy dependence of the optical potential predicted by the CEG07 folding model.

The channel-coupling effect is also analyzed in terms of the dynamical polarization potential (DPP).
The DPP drastically changes with the incident energy.
The real part of the DPP shows the transition from repulsion to attraction in the energy evolution, whereas its imaginary part shows the transition from negative to positive.
These transitions reflect the characteristic energy dependence of the complex coupling potential, which is clearly understood by the close relation between the real and imaginary parts of the DPP and the real and imaginary parts of the complex coupling potential.

We have found that inelastic cross sections at these incident energies are dominated by the imaginary part of the coupling potential, which also reflect the characteristic energy dependence of the real and imaginary parts of the coupling potential. 
This suggests that the measurement of the absolute magnitude of the inelastic cross sections at very forward angles at these incident energies will provide a crucial test for the validity of microscopic interaction models and removes the ambiguity of the strength of the imaginary coupling potential.

\section{Acknowledgment}
The authors acknowledge Professor K.~Ogata for valuable comments.
One of authors (T.F.) is supported by the Special Postdoctoral Researcher Program of RIKEN.



\end{document}